\documentclass[11pt,en]{elegantpaper}
\title{Rescattering effect on the measurement of $K^{\ast\notag0}$ spin alignment in heavy-ion collisions}
\author{Ziyang Li, Wangmei Zha\footnote{first@ustc.edu.cn}, and Zebo Tang\footnote{zbtang@ustc.edu.cn}}
\institute{State Key Laboratory of Particle Detection and Electronics, University of Science and Technology of China, Hefei 230026, China}
\date{}
\usepackage{graphicx}
\usepackage[caption=false,font=normalsize,labelfont=sf,textfont=sf]{subfig}
\usepackage{float}
\usepackage{lineno}
\RequirePackage{lineno}
\begin{document}

\maketitle

\begin{abstract}
Spin alignment of vector mesons in noncentral relativistic heavy ion collisions provides a novel probe of the global quark polarization and hadronization mechanism. In this paper, we discuss the spin alignment of a short-lived vector meson, namely $K^{\ast\notag0}$, arising from the hadronic rescattering process using the UrQMD model. This spin alignment is not related to global quark polarization but cannot be distinguished from those arising from global quark polarization in experiment. The spin alignment parameter $\rho_{00}$ is found to deviate from 1/3 by up to $-$0.008 (0.03) with respect to the reaction (production) plane. These deviations are much larger than the expected signal from all the theoretical models implementing the conventional global quark polarization mechanism as well as the current experimental precision, and should be considered seriously when comparing measurements with theoretical calculations. 

\end{abstract}

\section{Introduction}
Relativistic heavy-ion collisions create a hot, dense medium with a partonic degree of freedom, called the quark-gluon plasma (QGP). It provides unique opportunities to study quantum chromodynamics (QCD). In noncentral heavy-ion collisions, a large orbital angular momentum ($\approx 10^{4} ~\hbar$)~\cite{liang2005globally,Becattini:2007sr} and a strong magnetic field ($\approx 10^{14} ~T$)~\cite{Kharzeev:2007jp} are also expected. While the magnetic field is expected to be short-lived, part of the angular momentum is conserved and could be felt throughout the evolution of the system formed in the collision ~\cite{liang2005globally}. It is predicted that particles produced in heavy-ion collisions with large global angular momentum could be polarized globally in the direction of the orbital angular momentum due to spin-orbit coupling ~\cite{liang2005globally}. In 2017, the STAR Collaboration reported the discovery of global polarization for $\Lambda$ and $\bar{\Lambda}$ hyperons in Au+Au collisions~\cite{adamczyk2017global}. According to the flavor-spin wave function, the polarization of the $\Lambda$($\bar{\Lambda}$) hyperon is carried solely by the strange quark $s$($\bar{s}$), indicating the global polarization of the s($\bar{s}$) quark. \par 
The polarized quarks can also form polarized vector mesons such as $\phi(1020)$ and $K^{\ast}(892)$. These mesons originate predominantly from primordial production and are less affected by feed-down contribution compared to $\Lambda$ and $\bar{\Lambda}$ hyperons. Furthermore, as spin-1 particles, their daughter's polar angle distribution is an even function; thus, there is no local cancellation as associated with spin-1/2 hyperons when integrating over time and phase space.

The spin state of a vector meson can be described by a 3$\times$3 Hermitian spin density matrix $\rho$ with unit trace. The diagonal elements $\rho_{-1-1}$ and $\rho_{11}$ cannot be measured separately. Consequently, $\rho_{00}$ is the only independent element ~\cite{SCHILLING1970397}. It can be determined from the angular distribution of the decay products,
\begin{equation}\label{eq1}
\frac{dN}{d\cos\theta^*}\propto(1-\rho_{00}) + (3\rho_{00}-1)\cos^{2}\theta^{*},
\end{equation}
where $\theta^{*}$ is the angle between the polarization direction and the momentum direction of one of the decay daughters in the rest frame of the vector meson. $\rho_{00}$ is 1/3 in the absence of spin alignment, and a deviation of $\rho_{00}$ from 1/3 signals net spin alignment.

The ALICE Collaboration recently published the measurements of vector mesons $K^{\ast\notag0}$ and $\phi$ spin alignment in Pb+Pb collisions at 2.76 TeV~\cite{ALICE:2019aid}. $\rho_{00}$ values are found to decrease with decreasing transverse momentum ($p_T$), consisting of 1/3 at high $p_T$ but less than 1/3 at low $p_T$ ($p_T < 2~\text{GeV}/c$) at a level of 3$\sigma$ (2$\sigma$) for $K^{\ast\notag0}$ ($\phi$), respectively. The values at low $p_T$ are systematically smaller for $K^{\ast\notag0}$ than for $\phi$, and with respect to the production plane than that with respect to the event plane. The difference is on the order of a few percent, although comparable to the uncertainties. The deviation of $\rho_{00}$ from 1/3 at low $p_T$ for the vector mesons is qualitatively consistent with the expectation from theoretical models, which attribute it to a polarization of quarks in the presence of angular momentum in heavy-ion collisions and subsequent hadronization by the process of recombination. However, the measured spin alignment is unexpectedly large compared to the global polarization of $\Lambda$ hyperons~\cite{ALICE:2019onw}. This is therefore puzzling.

The STAR Collaboration measured the spin alignment for $K^{\ast\notag0}$ and $\phi$ in Au+Au collisions at $\sqrt{s_{\mathrm{NN}}}$ from 11.5$-$200 GeV~\cite{STAR:2022spinalignment}. $\rho_{00}$ for $K^{\ast\notag0}$ is largely consistent with 1/3 [$0.3356 \pm 0.0034 ~\textrm{(stat.)} \pm 0.0043 ~\textrm{(syst.)}$ at $\sqrt{s_{\mathrm{NN}}}\leq 54.4~\textrm{GeV}$], while $\rho_{00}$ for $\phi$ is above 1/3 with a significance of $7.4\sigma$ [$0.3512 \pm 0.0017 ~\textrm{(stat.)} \pm 0.0017 ~\textrm{(syst.)}$ at $\sqrt{s_{\mathrm{NN}}}\leq 62~\textrm{GeV}$]. The large difference between these two vector mesons cannot be described by any theoretical models with conventional global quark polarization mechanisms~\cite{PhysRevD.105.L011901,PhysRevC.98.014910,adamczyk2017global,PhysRevD.104.076016,PhysRevD.102.056013,XIA2021136325,PhysRevC.97.034917,liang2005spin,sheng2020can}. A recent theoretical model based on $\phi$-meson vector field coupling to strange quark can qualitatively describe the difference, but more developments are needed for quantitative description~\cite{STAR:2022spinalignment,sheng2020can}. On the other hand, it is important to survey the other possible sources contributing to the difference between $K^{\ast\notag0}$ and $\phi$ mesons spin alignment.

The lifetime of $K^{\ast\notag0}$ is approximately 4 fm/$c$, which is approximately ten times shorter than the $\phi$ lifetime and comparable to the time between chemical freeze-out and kinetic freeze-out, making it sensitive to hadronic interactions. In experiment, the vector mesons are usually reconstructed via their strong decay to hadrons. If a $K^{\ast\notag0}$ decays before kinetic freeze-out, the decay daughters are subject to rescattering by hadrons via elastic or inelastic collisions. If one or more daughter particles are rescattered, either its momentum is altered or completely absorbed, the parent resonance cannot be reconstructed experimentally~\cite{Bleicher:2003ij,Knospe:2015nva,Steinheimer:2017vju,Bleicher:2002dm}. The measurements of centrality and $p_T$ dependence of $K^{\ast\notag0}$ yield in heavy-ion collisions show strong evidence~\cite{STAR:2004bgh,PhysRevC.84.034909,ALICE_Kstar_rescattering} of the significant rescattering effect on $K^{\ast\notag0}$ production.

The rescattering probability depends on the particle density profile, which is related to the geometry of the medium. In noncentral heavy ion collisions, the rescattering probability as a function of azimuthal angle may be an anisotropic distribution due to the initial geometry of the collisions. Since $\rho_{00}$ of vector mesons are measured via the angular distribution of the daughter particle with respect to the event plane or production plane, it may be distorted by the rescattering effect for reconstructable resonances. Consequently, the rescattering effect could produce artificial spin alignment for resonances, such as $K^{\ast\notag0}$. 

In this paper, the spin alignment of $K^{\ast\notag0}$ arising from hadronic rescattering in Au+Au collisions at $\sqrt{s_{\mathrm{NN}}} = 200~\textrm{GeV}$ is studied with the ultrarelativistic quantum molecular dynamics (UrQMD) model. The analysis method is introduced in Sec.~\ref{sec:method}. Seciton~\ref{sec:results} includes the results and discussions. The paper is summarized in Sec.~\ref{sec:summary}.

\section{Analysis Method}\label{sec:method}


\begin{figure}[htbp]
	\centering
	\includegraphics[width=0.5\linewidth]{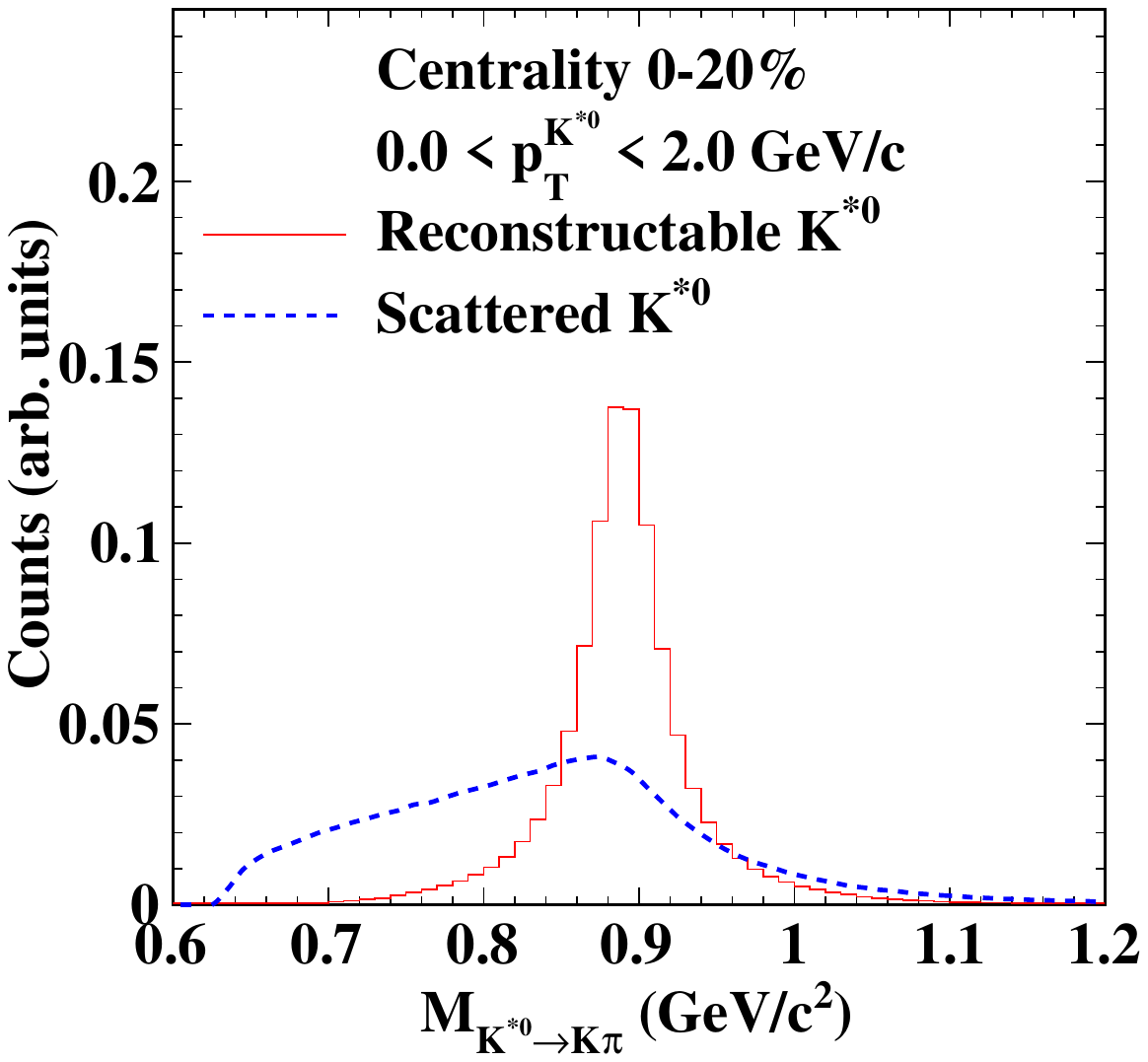}
	\caption{Invariant mass distribution of $K^{*0}\rightarrow K\pi$ for reconstructable $K^{*0}$ (red solid line) and un-reconstructable $K^{*0}$ (blue dot-dashed line) in 0-20$\%$ centrality. The un-reconstructable $K^{*0}$ is those with one or two daughters experiencing at least one elastic scattering. The distributions are scaled by their integral.
	\label{fig:invmass}}
\end{figure}

\begin{figure}[htbp]
\centering
\captionsetup[subfloat]{labelsep=none,format=plain,labelformat=empty}
\subfloat[]{
	\begin{minipage}[t]{0.45\linewidth}
		\centering
		\includegraphics[width=3in]{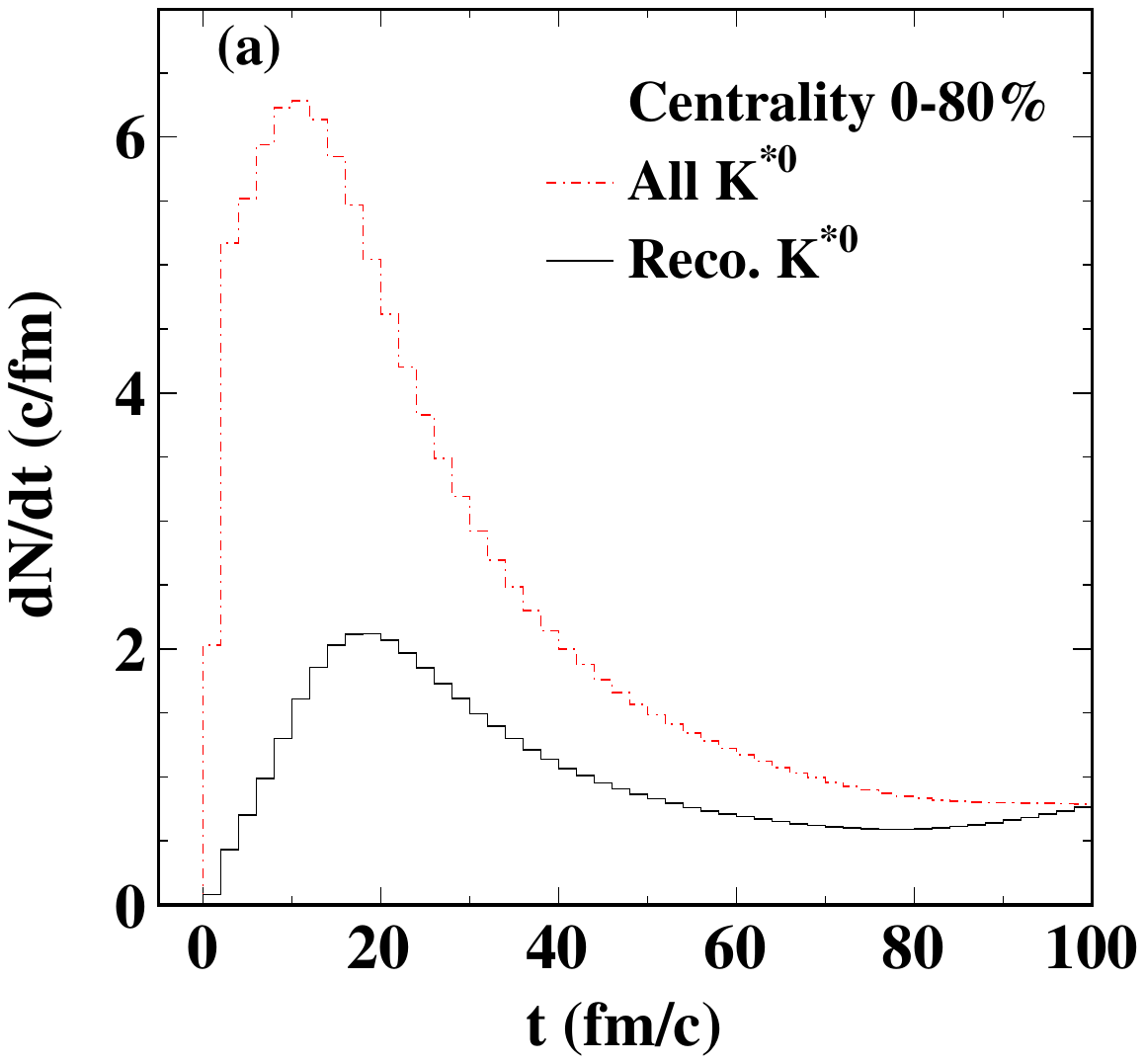}
		
	\end{minipage}
 }
\subfloat[]{
	\begin{minipage}[t]{0.45\linewidth}
		\centering
		\includegraphics[width=3in]{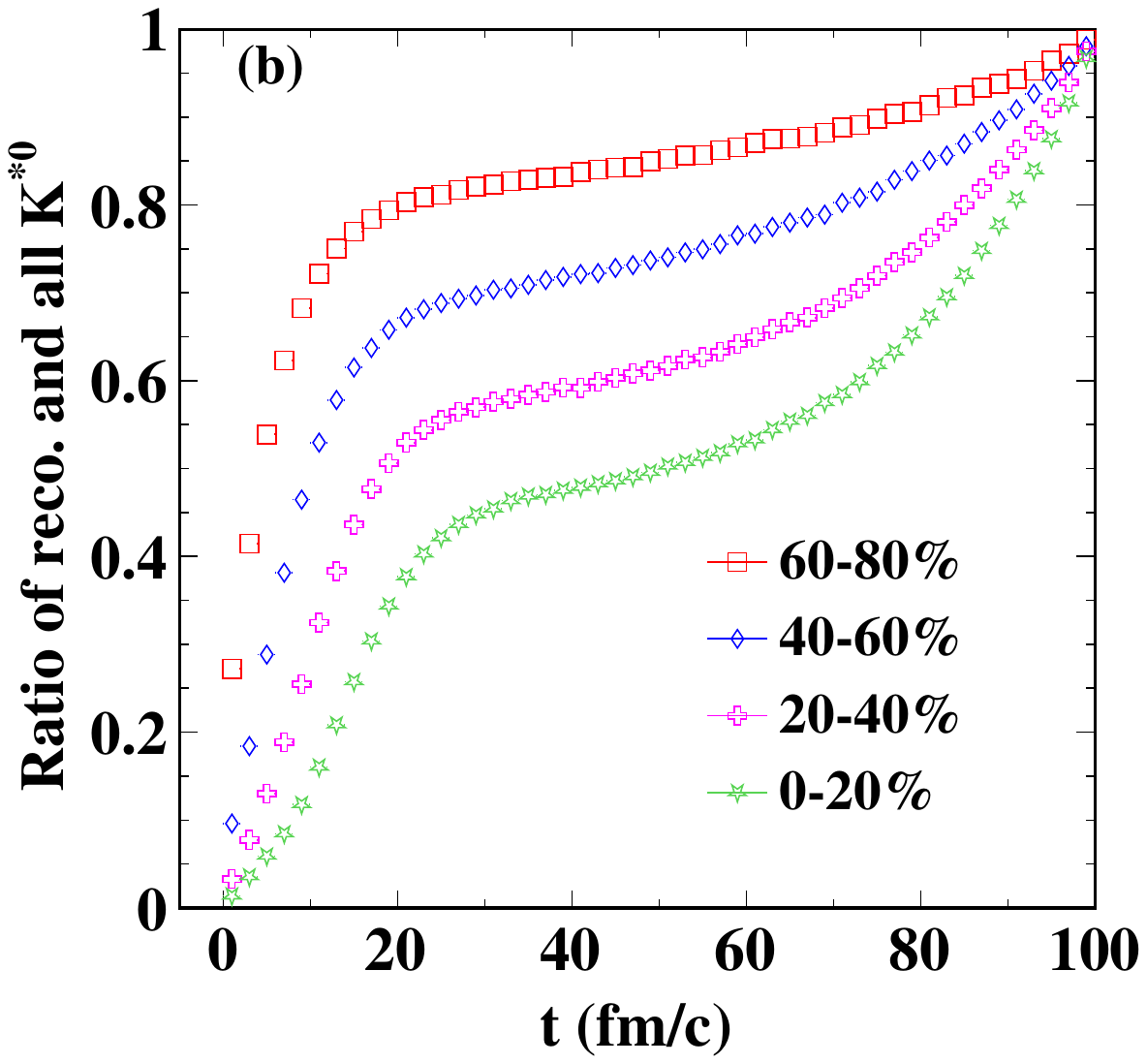}
	\end{minipage}
}
\centering
\caption{(a) The yield of all (red line) and reconstructable (black line) $K^{\ast0}$ as a function of the time when they decayed in Au+Au collisions at $\sqrt{s_\mathrm{NN}}$ = 200 GeV for 0-80$\%$ centrality from the UrQMD. (b) The ratio of reconstructable and all $K^{\ast}$ as a function of the time for various centralities.
\label{fig:time distribution}}

\end{figure}

\begin{figure}[htbp]
	\centering
	\includegraphics[width=0.5\linewidth]{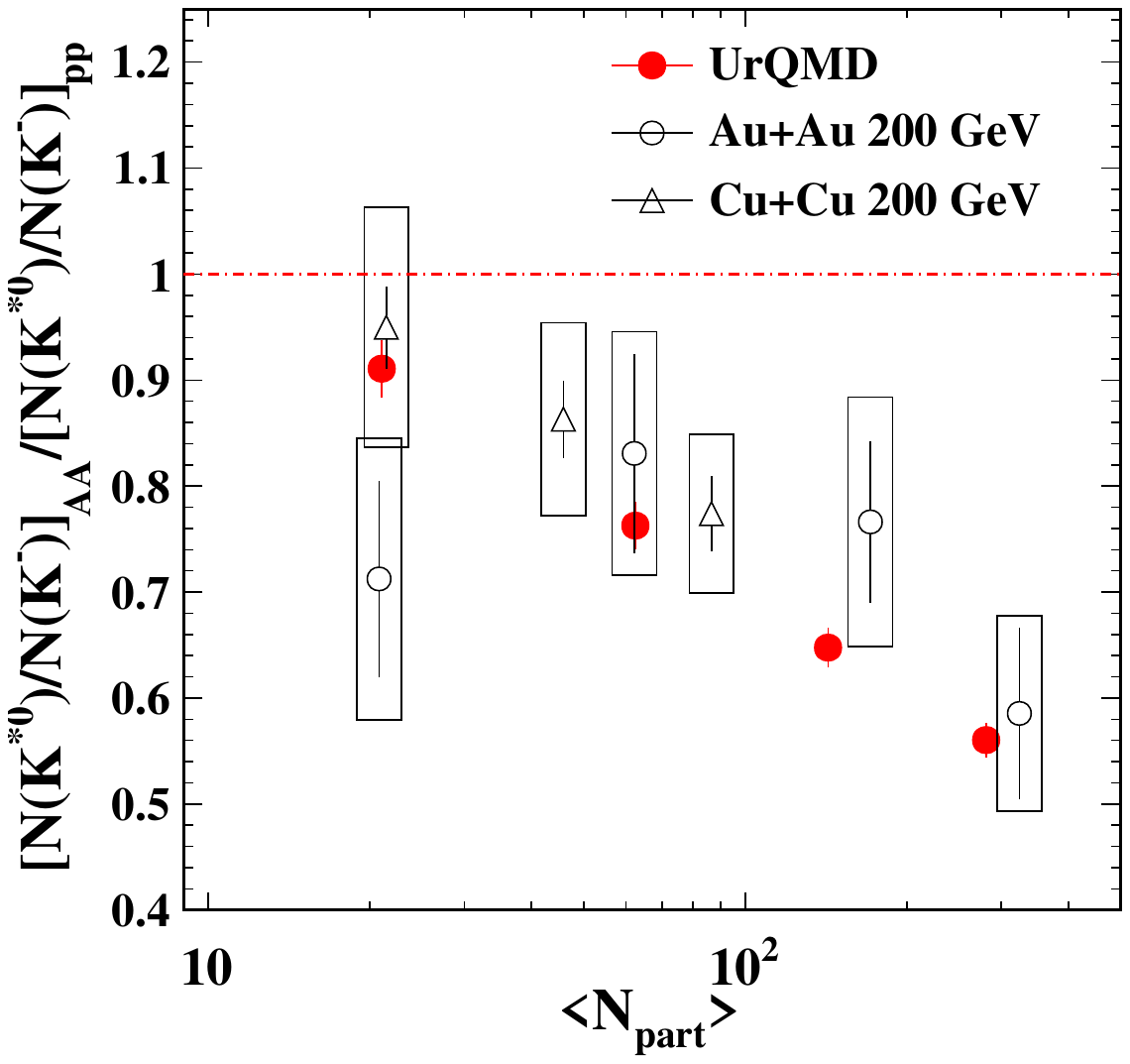}
	\caption{The yield ratio of reconstructable $K^{\ast0}$ over $K^-$ in Au+Au collisions from UrQMD divided by that measured in $p + p$ collisions as a function of $\langle N_{\textrm{part}} \rangle$, compared with measurements in Au + Au and Cu + Cu collisions~\cite{PhysRevC.84.034909}. The center-of-mass energy is $\sqrt{s_\mathrm{NN}}$=200 GeV.
	\label{fig:doubleratio}}
\end{figure}

The spin alignment of $K^{\ast\notag0}$ arising from hadronic rescattering is investigated with the UrQMD model. UrQMD is a microscopic transport approach based on the covariant propagation of hadrons and strings. All cross-sections are calculated by the principle of detailed balance or are fitted to available data. The initial orbital angular momentum are not explicitly propagated in UrQMD therefore no spin alignment for initially produced vector mesons in UrQMD. Any deviation of $\rho_{00}$ from 1/3 indicates the influences on polarization from hadronic interactions. One important feature of UrQMD is that it has the option to save all the collision history of particles during the evolution of the system, making it an ideal tool to study the hadronic rescattering effect.

As mentioned above, $K^{\ast\notag0}$ is usually reconstructed via the decay of $K^{\ast\notag0}\rightarrow K\pi$. In a heavy-ion collision such as Au+Au collision at $\sqrt{s_{\mathrm{NN}}}$ = 200 GeV, hundreds or even thousands of pions and kaons are produced. In experiment, to reconstruct $K^{\ast\notag0}\rightarrow K\pi$, one needs to pair all the pion candidates with all the kaon candidates in the same event and calculate their invariant masses. With this method, the decay of $K^{\ast\notag0}\rightarrow K\pi$ can be reconstructed, together with a huge random combinatorial background. Although the random combinatorial background can be reproduced with the like-sign or mixed-event technique and statistically subtracted~\cite{STAR:2004bgh}, 
the measurement of $K^{\ast\notag0}$ is very challenging as the typical signal-to-background ratio is on the order of 1/100. In many heavy-ion collision event generators, only the final particles are saved. One needs to reconstruct $K^{\ast\notag0}$ in the same way as in experiment. Alternatively, one can turn off the decay of $K^{\ast\notag0}$ and directly access the information of $K^{\ast\notag0}$ from the output file. However, in this case, $K^{\ast\notag0}$ does not suffer from the rescattering effect, which is the main focus of this paper. \par

Thanks to the collision history file of UrQMD, one can easily study the rescattering effect of resonance. The history file contains each binary interaction, resonance decay, and string excitation that occurred in the course of the heavy-ion reaction. First, we find all the $K^{\ast\notag0}\rightarrow K^\pm\pi^\mp$ decay with $|y(K^{\ast\notag0})|<0.5$ from the collision history file. We then follow the collision history of the decay daughters. If any of the decay daughters participates in any (elastic or inelastic) collision with other hadrons, the $K^{\ast\notag0}$ is categorized as ``un-reconstructable''. If neither of the two decay daughters participates in any collision with other hadrons, the $K^{\ast\notag0}$ is categorized as ``reconstructable''. Since it has been found that a finite $\eta$ acceptance will lead to an increase in the observed $\rho_{00}$~\cite{LAN2018319}, we apply no $\eta$ acceptance cut to decay daughters. Figure~\ref{fig:invmass} shows the invariant mass distribution of $K^{\ast\notag0}\rightarrow K^\pm\pi^\mp$ in 0-20$\%$ centrality. The solid red line depicts the invariant mass distribution of $\pi$ and K from the decay of reconstructable $K^{\ast\notag0}$. It follows the Breit-Wigner distribution. The dashed blue line depicts the invariant mass distribution of $\pi$ and K from $K^{\ast\notag0}$ with at least one of them participating in an elastic collision with other hadrons and the momentum after the collision is used in the invariant mass calculation. The distribution is significantly broadened. In reality, the distribution should be even broader as the decay daughter measured by detectors may have experienced more than one elastic collision. These $K^{\ast\notag0}$ cannot be reconstructed in the presence of a large random combinatorial background and significantly correlated background in heavy ion collisions. Not presented in Fig.~\ref{fig:invmass} is the $K^{\ast\notag0}$ with one or two decay daughters absorbed in the following collisions. It is completely lost. Figure~\ref{fig:time distribution}(a) shows the number of all and reconstructable $K^{\ast\notag0}$ as a function of the time when they decay to K and $\pi$ in 0-80\% Au+Au collisions at $\sqrt{s_{\mathrm{NN}}}$ = 200 GeV. The mean of the time distribution for both all and reconstructable $K^{\ast}$ is between 10 and 30 fm/$c$, which means that most $K^{\ast\notag0}$ decays during this time. Figure~\ref{fig:time distribution}(b) shows the ratio of reconstructable over all $K^{\ast\notag0}$ as a function of time in various centrality classes. It increases with increasing time and toward peripheral collisions, showing strong evidence of a significant rescattering effect implemented in the UrQMD model, especially at low-$p_T$ and in central collisions. Figure~\ref{fig:doubleratio} shows the $K^{\ast\notag0}/K^{-}$ ratio in UrQMD normalized by the ratio measured in $p + p$ collisions at $\sqrt{s_{\mathrm{NN}}}$ = 200 GeV~\cite{PhysRevC.84.034909}. The double ratio decreases with increasing system size, consistent with the experimental results in Au + Au and Cu + Cu collisions~\cite{PhysRevC.84.034909}. This comparison verifies that the rescattering processes implemented in the UrQMD model are reasonably good.\par

\begin{figure}[htp]
\centering
	\begin{minipage}[t]{0.4\linewidth}
		\centering
		\includegraphics[width=1.0\linewidth]{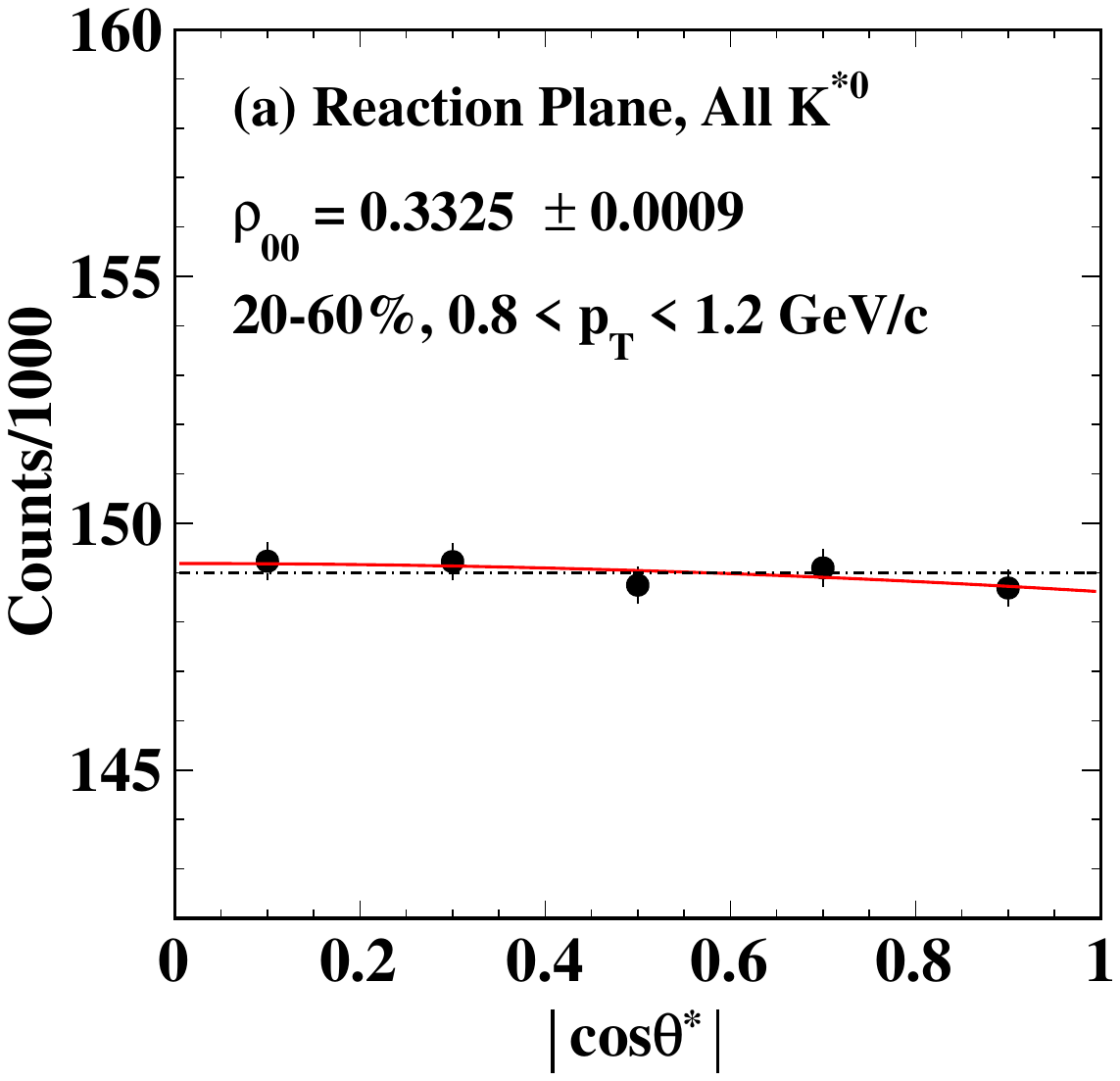}
	\end{minipage}
	\begin{minipage}[t]{0.4\linewidth}
		\centering
		\includegraphics[width=1.0\linewidth]{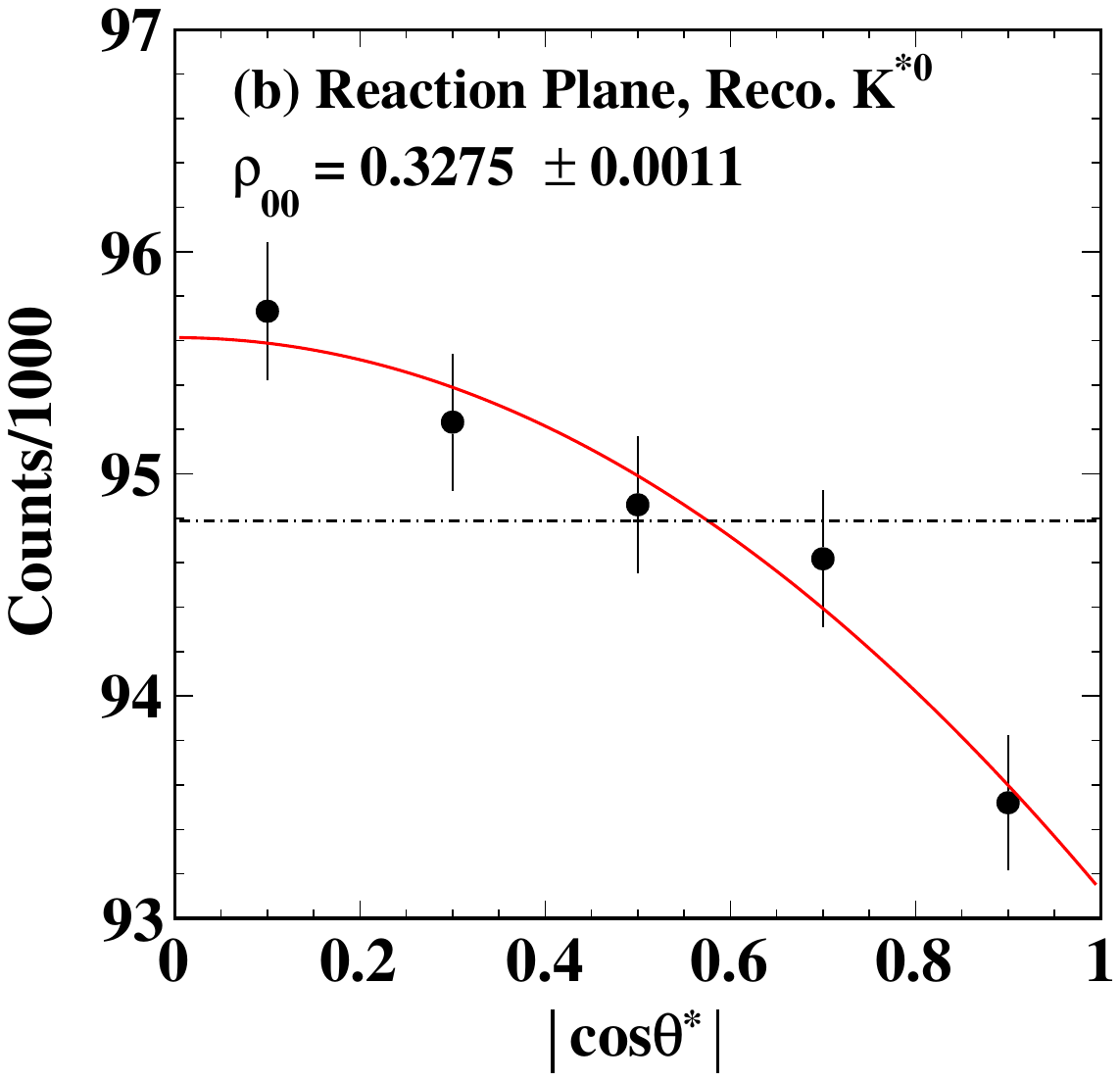}
	\end{minipage}
	\\
	\begin{minipage}[t]{0.4\linewidth}
		\centering
		\includegraphics[width=1.0\linewidth]{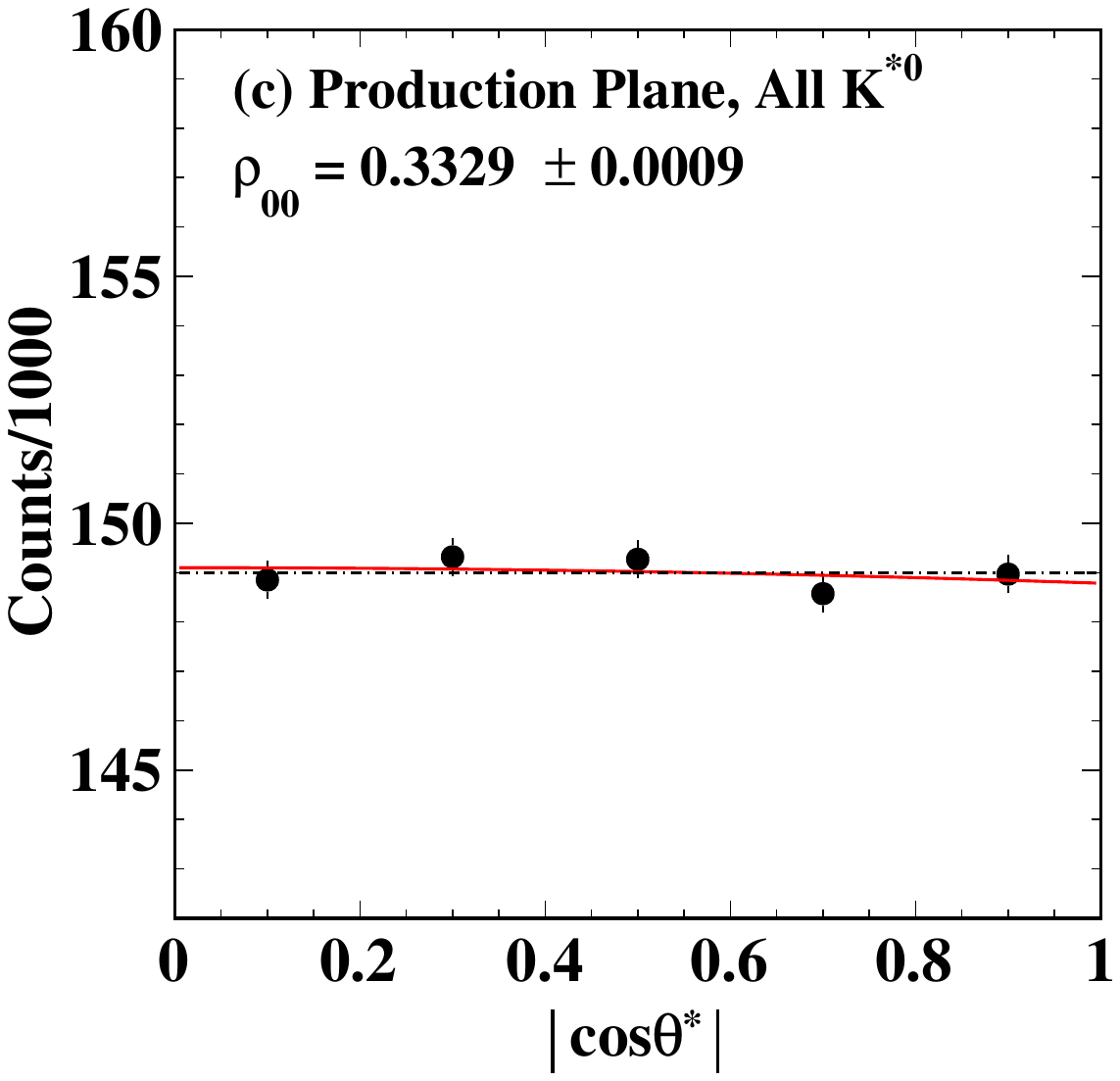}
	\end{minipage}
	\begin{minipage}[t]{0.4\linewidth}
		\centering
		\includegraphics[width=1.0\linewidth]{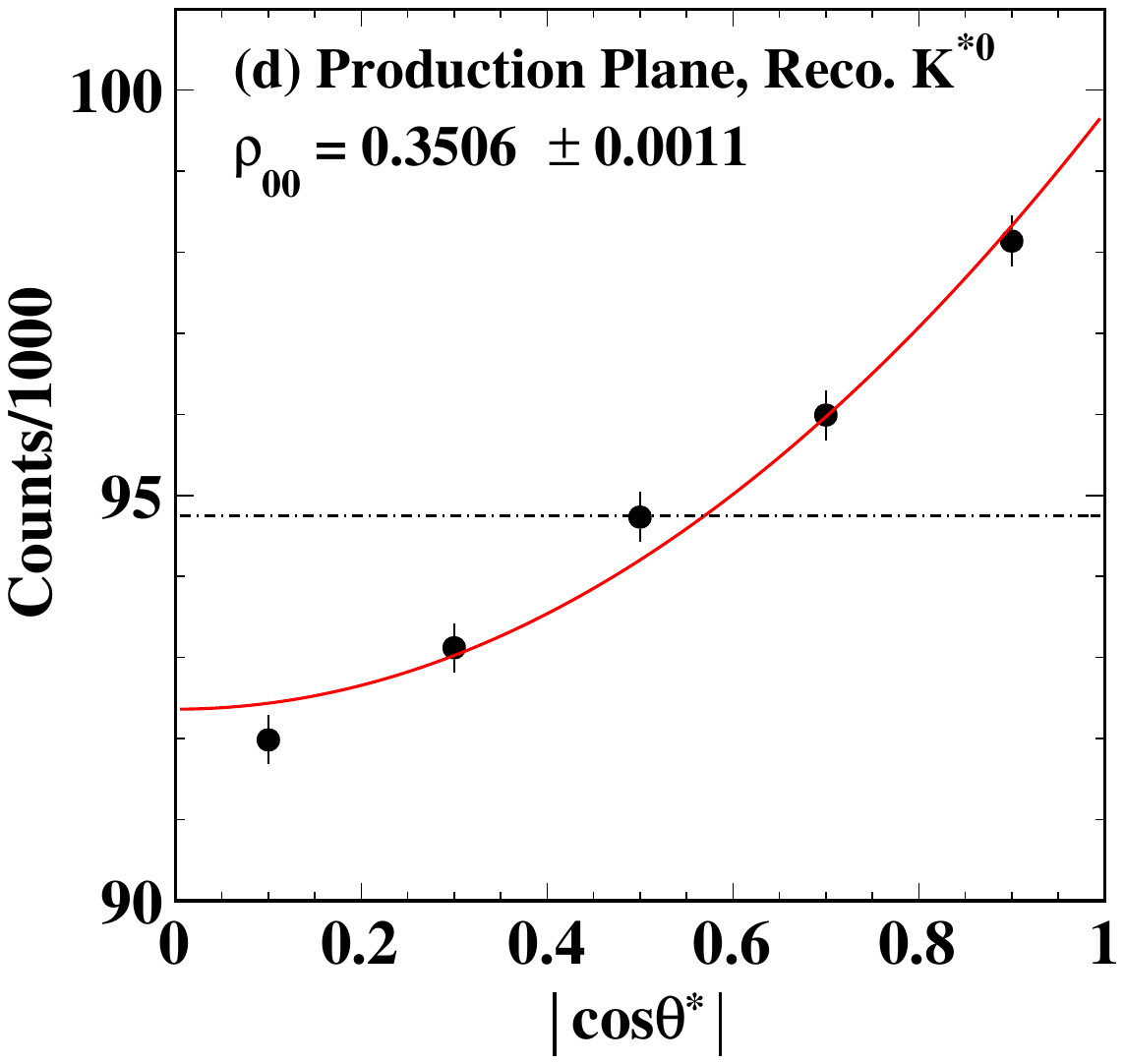}
	\end{minipage}
\centering
\caption{The $\cos\theta^{*}$ distributions for (a), (c) all $K^{\ast}$ and (b), (d) reconstructable $K^{\ast}$ for 0.8 < $p_{T}$ < 1.2 GeV/c in 20-60$\%$ centrality with respect to the reaction plane and production plane, respectively. The solid lines represent the fit to Eq.~\ref{eq1} and the dot-dashed lines represent the fit to a constant.  
\label{fig:AbsCostheta_RP}}
\end{figure}

Spin alignment of vector mesons is usually studied with respect to reaction plane or production plane in literature~\cite{PhysRevD.105.L011901,PhysRevC.98.014910,adamczyk2017global,PhysRevD.104.076016,PhysRevD.102.056013,XIA2021136325,PhysRevC.97.034917,liang2005spin,sheng2020can,SCHILLING1970397,ALICE_Kstar_rescattering,STAR:2022spinalignment,ALICE:2019aid}. In this study, the spin alignment parameter $\rho_{00}$ of all and reconstructable $K^{\ast\notag0}$ is extracted with respect to the reaction plane and production plane, respectively, according to Eq.~(\ref{eq1}). The reaction plane is defined by the cross product of the beam momentum and the impact parameter, while the production plane is defined by the cross product of the beam momentum and $K^{\ast0}$ momentum. The polarization direction is along the direction of the planes. Figure ~\ref{fig:AbsCostheta_RP} shows an example of the $\cos\theta^*$ distribution with respect to the reaction plane and production plane for all [Figs. 4(a), 4(c)] and reconstructable [Figs. 4(b), 4(d)] $K^{\ast0}$ at $0.8<p_T<1.2~\textrm{GeV}/c$ in 20-60\% Au+Au collisions at $\sqrt{s_{\mathrm{NN}}}$ = 200 GeV. The solid lines represent the fit to Eq.~(\ref{eq1}).

\section{Results and Discussions}\label{sec:results}
\subsection{Spin alignment with respect to the reaction plane}

\begin{figure}[htbp]
	\centering
	\captionsetup[subfloat]{labelsep=none,format=plain,labelformat=empty}
	\subfloat[]{
		\begin{minipage}[t]{0.45\linewidth}
			\centering
			\includegraphics[width=3in]{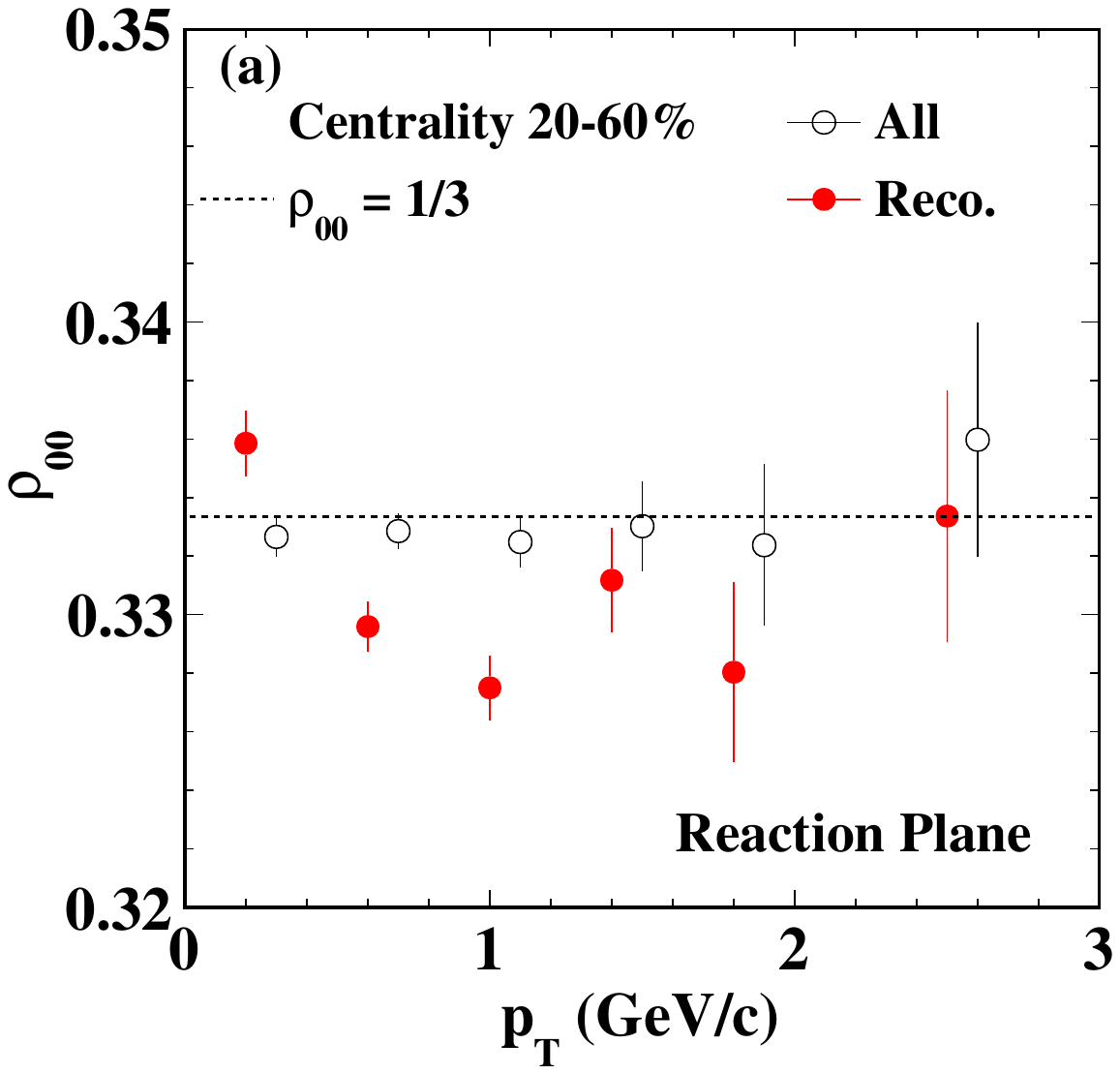}
		\end{minipage}
	}
	\subfloat[]{
		\begin{minipage}[t]{0.45\linewidth}
			\centering
			\includegraphics[width=3in]{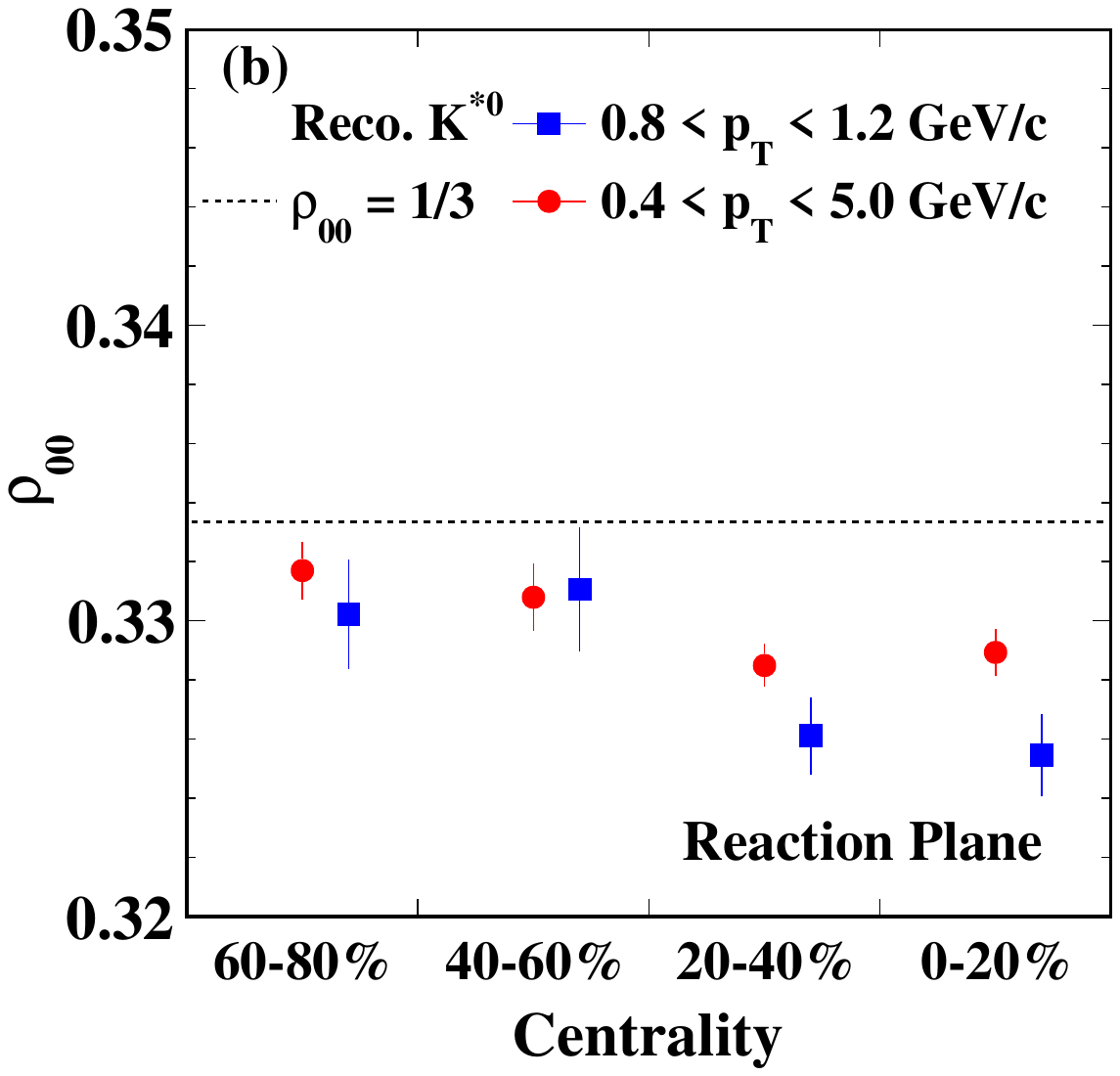}
		\end{minipage}
	}
	\centering
	\caption{ $\rho_{00}$ with respect to the reaction plane of all (open) and reconstructable (filled) $K^{\ast0}$ as a function of $p_T$ in 20-60$\%$ centrality (a) and that of reconstructable $K^{\ast0}$ as a function of centrality in two different $p_T$ intervals (b) in Au+Au collisions at  $\sqrt{s_{\mathrm{NN}}}$ = 200 GeV. 
	\label{fig:AbsCosthetavsptandcent_RP}}
\end{figure}

We first investigated the spin alignment of all $K^{\ast0}$. As shown in Fig.~\ref{fig:AbsCostheta_RP}(a), the $\cos\theta^*$ distribution is flat for all $K^{\ast0}$ in 20-60\% Au+Au collisions and the extracted $\rho_{00}=0.3325 \pm 0.0009$ is consistent with 1/3 within uncertainties. The $p_T$ dependence is shown in Fig.~\ref{fig:AbsCosthetavsptandcent_RP}(a) as open circles. No spin alignment for all $K^{\ast0}$ is observed, which proves that the initially produced $K^{\ast0}$ has no spin alignment in the UrQMD model as expected.

We then studied the rescattering effect on the $\rho_{00}$ measurement. Figure~\ref{fig:AbsCostheta_RP}(b) shows that there is less reconstructable $K^{\ast0}$ toward $|\cos\theta^*|=1$ or $\theta^*=0~\textrm{or }\pi$. This is because there are more particles out of the reaction plane than in the reaction plane in noncentral heavy ion collisions. The decay daughters of $K^{\ast0}$ have a larger rescattering probability along the polarization direction than perpendicular to the polarization direction. Consequently, there should be less reconstructable $K^{\ast0}$ at $|\cos\theta^*| \approx 1$ than $|\cos\theta^*| \approx 0$. It is noted that the picture is oversimplified, the angular distribution should be diluted by the momentum of $K^{\ast0}$ and the rescattering probability depends not only on the hadron density profile but also on the momentum distribution. The UrQMD model takes all these factors into account. This results in $\rho_{00}$ in this $p_T$ interval and centrality class being $0.3275 \pm 0.0011$, which is lower than 1/3 by 0.0058 ($>5\sigma$). Figure~\ref{fig:AbsCosthetavsptandcent_RP} shows its $p_T$ and centrality dependence. The deviation of reconstructable $K^{\ast0}$ $\rho_{00}$ from 1/3 is more significant in intermediate $p_T$ region and (semi-)central collisions, reaches as large as $-$0.008 at $0.8<p_T<1.2~\textrm{GeV}/c$ and 0-20\% centrality. 

Such a magnitude of deviation is significant compared with various theoretical sources. In previous theoretical calculations, the sources include the vorticity, electromagnetic fields, vector meson fields, fragmentation of polarized quarks, axial or helicity charge, and turbulent color fields. Since the different kinds of meson fields do not have a large correlation in space, the contributions from vector meson fields are absent for $K^{\ast0}$. The spin alignment of $K^{\ast0}$ is dominated by the vorticity field. The most obvious source would be the electric part of the vorticity tensor, which gives a negative contribution on the order of $10^{-4}$~\cite{PhysRevD.102.056013,sheng2020can}. The deviation of $\rho_{00}$ from the rescattering effect we measured is about an order of magnitude higher than this most obvious source. Then the magnetic part of the vorticity tensor gives a negative contribution, while the electric field gives a positive contribution, and the fragmentation of polarized quarks can give either a positive or negative contribution. These three sources are at the level of $10^{-5}$~\cite{sheng2020can,liang2005spin}. Helicity polarization gives a negative contribution at all centralities~\cite{PhysRevD.104.076016}. The exception from turbulent color fields is $\rho_{00}(K^{*0})<\rho_{00}(\phi)<1/3$~\cite{PhysRevD.105.L011901}.

\begin{figure}[htbp]
	\centering
	\includegraphics[width=0.45\linewidth]{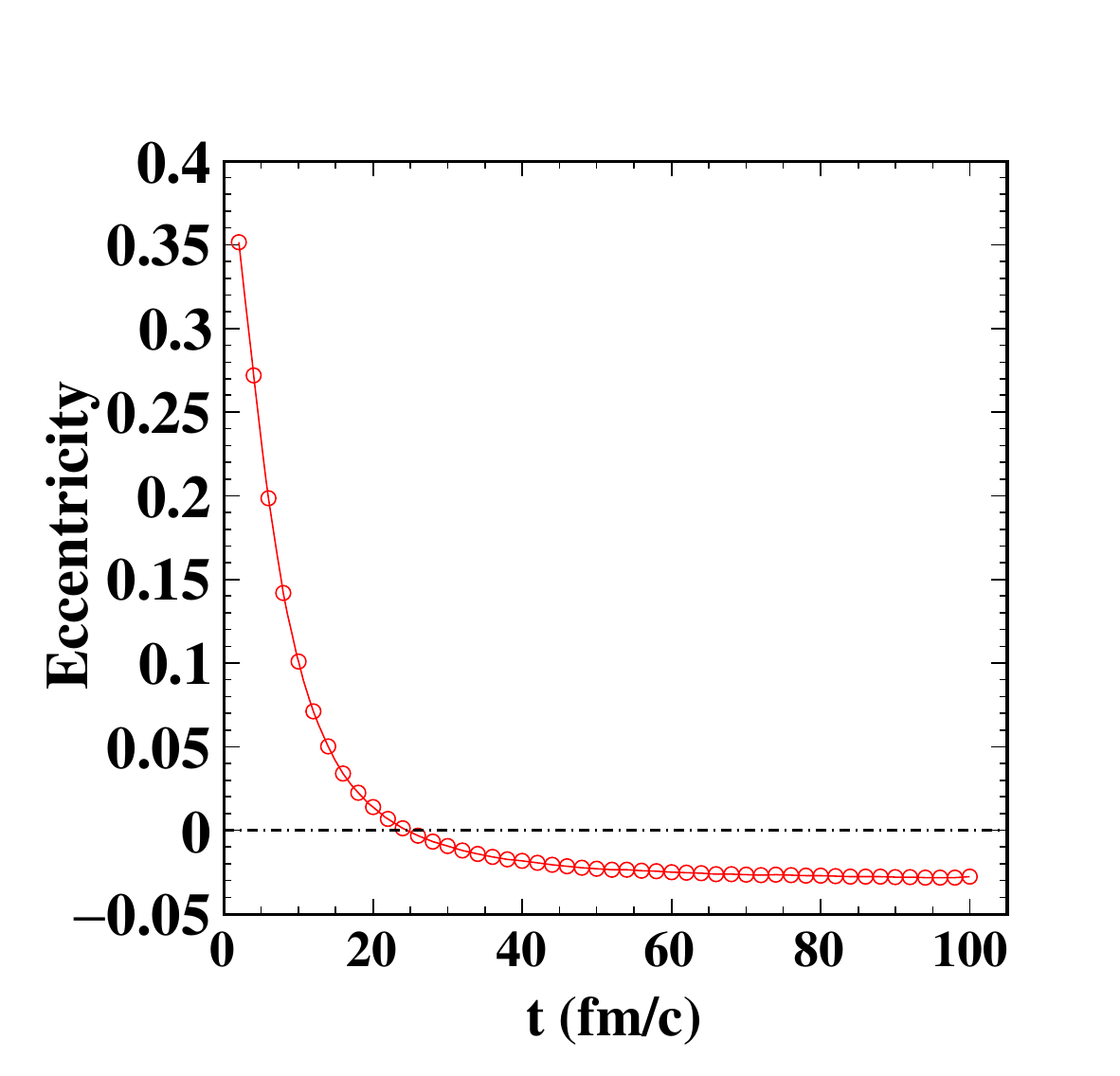}
	\caption{Eccentricity of $\pi^{\pm}$ as a function of the evolution time in 20-60$\%$ centrality
	\label{fig:eccentricity}}
\end{figure}

\begin{figure}[htbp]
\centering
\captionsetup[subfloat]{labelsep=none,format=plain,labelformat=empty}
\subfloat[]{
	\begin{minipage}[t]{0.4\linewidth}
		\centering
		\includegraphics[width=1.0\linewidth]{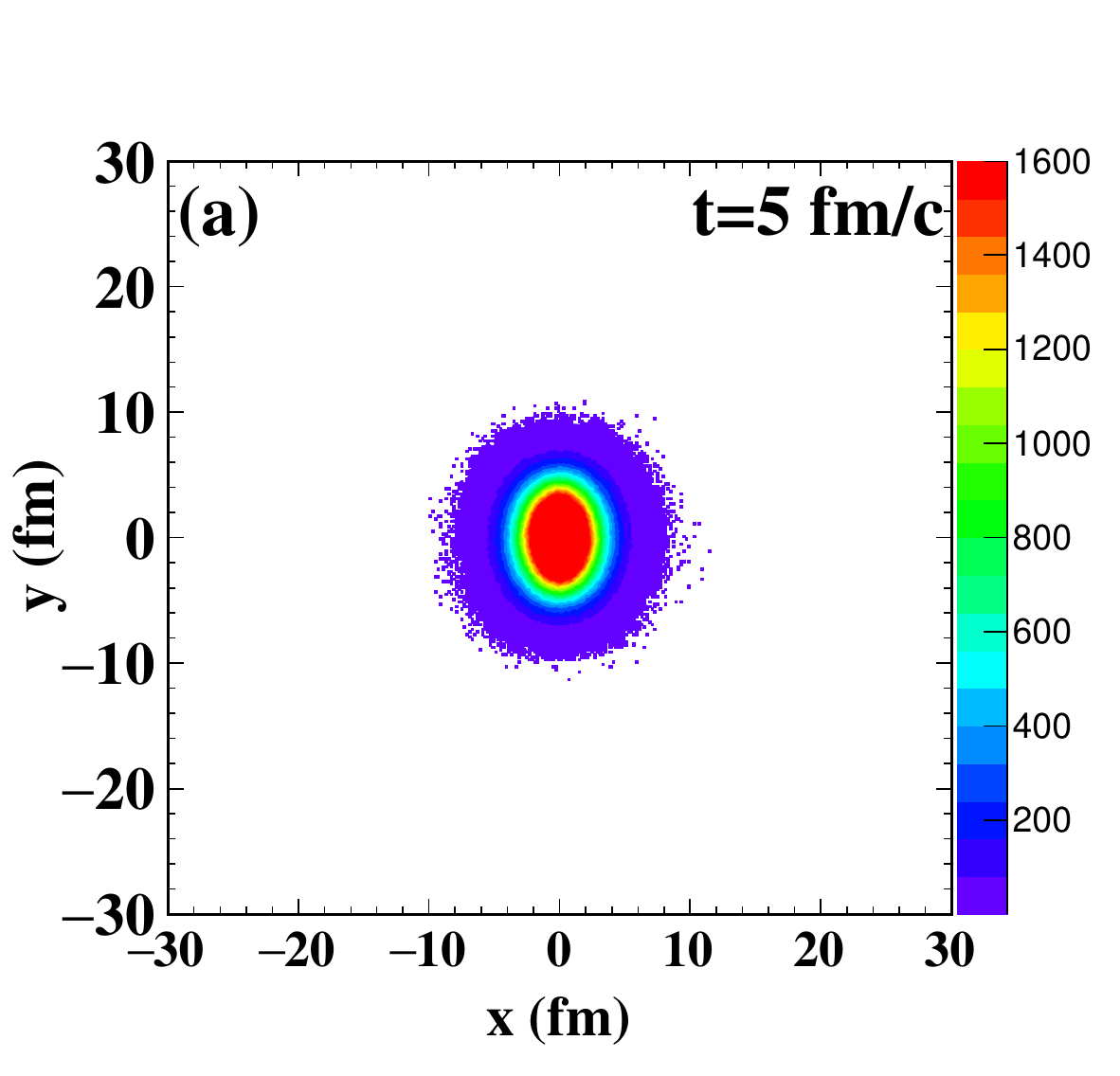}
	\end{minipage}
}
\subfloat[]{
	\begin{minipage}[t]{0.4\linewidth}
		\centering
		\includegraphics[width=1.0\linewidth]{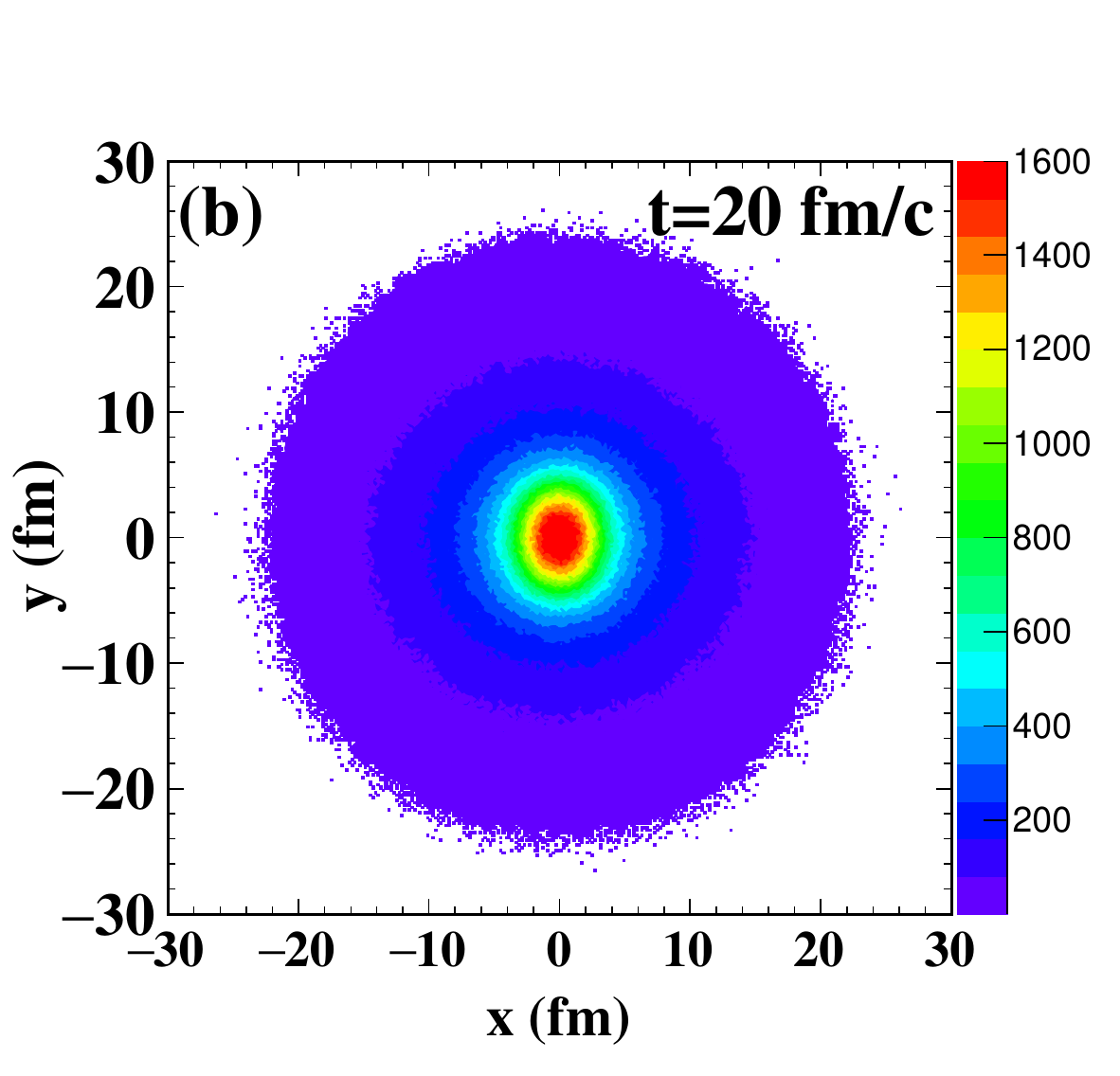}
	\end{minipage}
}
\centering
\caption{$\pi^{\pm}$ distribution in the transverse plane at (a) t = 5 fm/c and (b) 20 fm/c.
\label{fig:outtime}}
\end{figure}

Although the deviation of $\rho_{00}$ of $K^{\ast0}$ due to hadronic rescattering is significantly large compared to various theoretical calculations, it is much smaller compared to the naive estimation based on the initial eccentricity of the collisions or the anisotropy ($v_2$) of the final hadrons. Since the $\pi\pi$ cross section is much higher than that of $K\pi$, the rescattering effect of $K^{\ast0}$ is dominated by the scattering of the $\pi$ daughter with $\pi$s in the system. To better understand the behaviors of $K^{\ast0}$ $\rho_{00}$, we calculate the eccentricity of charged $\pi$ in coordinate space as a function of the evolution time in 20-60$\%$ centrality, as shown in Fig.~\ref{fig:eccentricity}. The eccentricity is defined as
\begin{equation}
\frac{\Sigma y^{2}-\Sigma x^{2}}{\Sigma y^{2}+\Sigma x^{2}},
\end{equation}
where $x$ is the position in the direction of the impact parameter, and $y$ is the position in the direction of the reaction plane. Eccentricity tends to be zero or even negative with time, which means the anisotropy of the medium decreases with the fireball evolution. Figure~\ref{fig:outtime} shows charged $\pi$ distributions in the transverse plane. The distribution is still an ellipse at 5 fm/$c$ but close to a circle at 20 fm/$c$. According to Fig.~\ref{fig:time distribution}, at the time when most $K^{\ast0}$ decays, the eccentricity of the medium is already small, and the loss of reconstructable $K^{\ast0}$ caused by rescattering is not so significantly different along different directions.  Note that the $v_2$ from UrQMD in cascade mode is smaller than data. The influence of spin alignment of $K^{\ast0}$ due to hadronic rescattering effect is expected to be even larger in real data than what is shown in Fig.~\ref{fig:AbsCosthetavsptandcent_RP}. \par

\subsection{Spin alignment with respect to the production plane}

\begin{figure}[htbp]
	\centering
	\captionsetup[subfloat]{labelsep=none,format=plain,labelformat=empty}
	\subfloat[]{
	\begin{minipage}[t]{0.45\linewidth}
		\centering
		\includegraphics[width=3in]{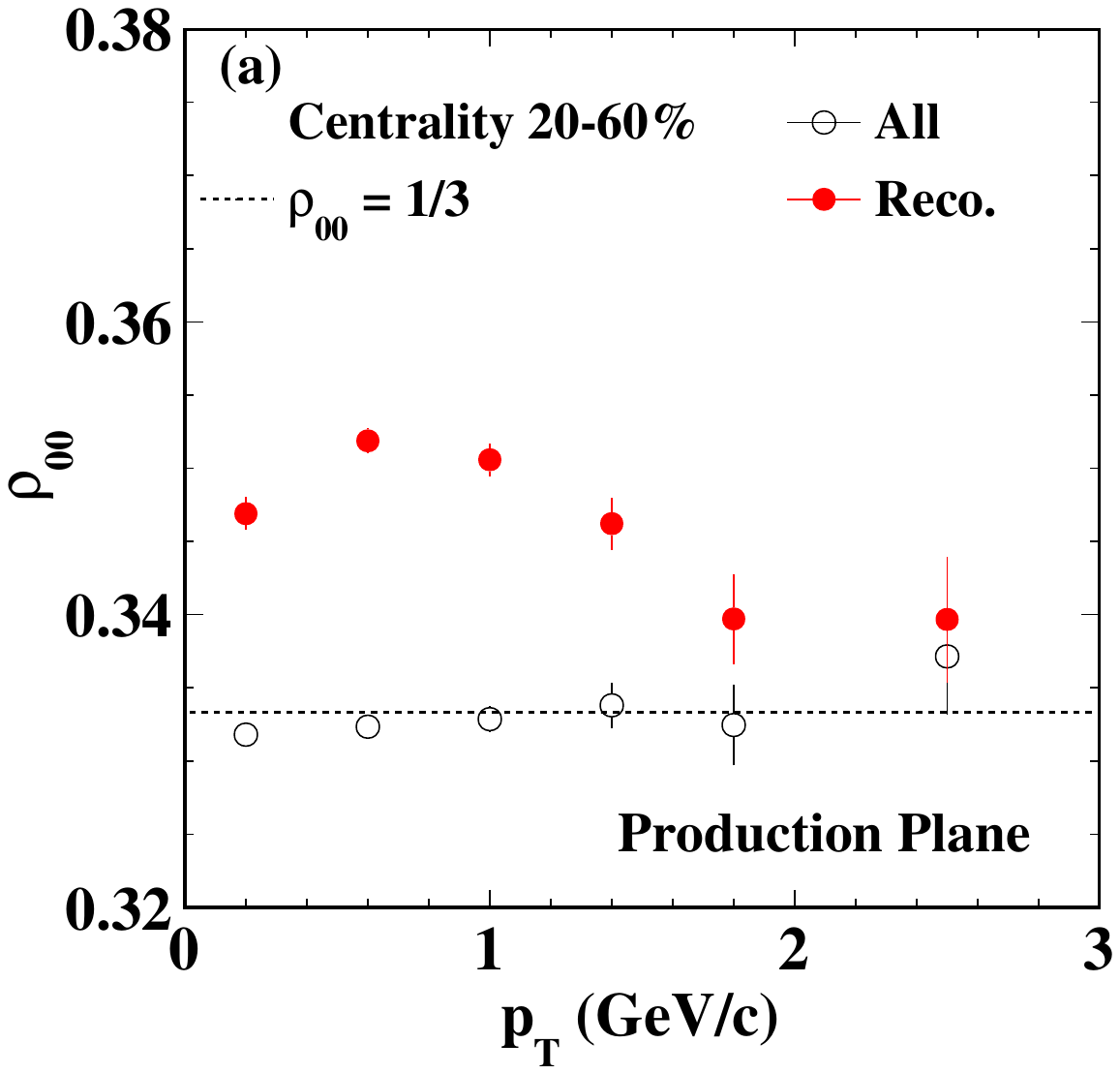}
	\end{minipage}
}
	\subfloat[]{
	\begin{minipage}[t]{0.45\linewidth}
		\centering
		\includegraphics[width=3in]{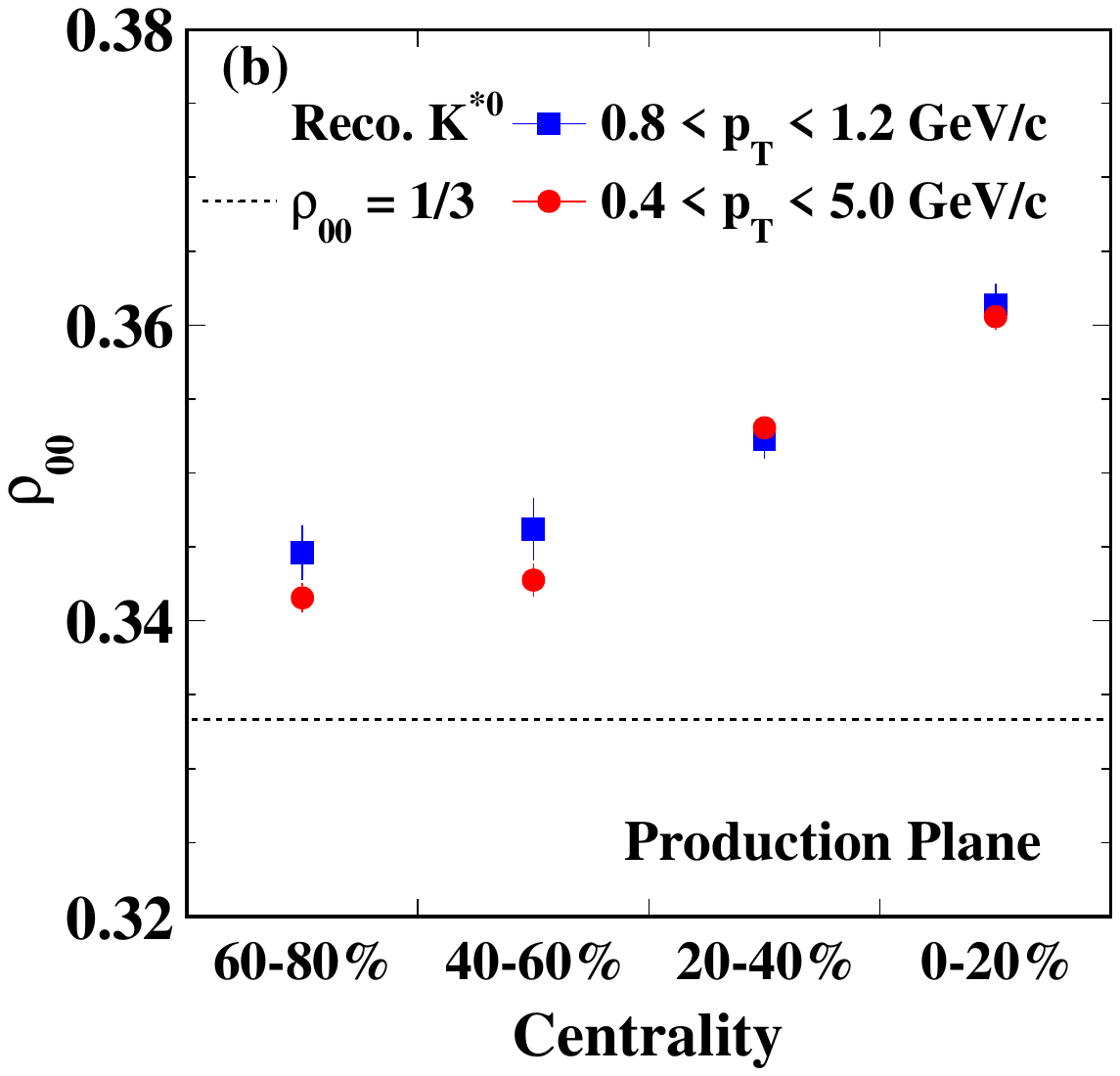}
	\end{minipage}
}
	\centering
	\caption{(a) $\rho_{00}$ with respect to the production plane of all and reconstructable $K^{\ast0}$ as a function of $p_T$ in 20-60$\%$ centrality and (b) that of reconstructable $K^{\ast0}$ as a function of centrality in two different $p_T$ intervals in Au+Au collisions at  $\sqrt{s_{\mathrm{NN}}}$ = 200 GeV.
	\label{fig:AbsCosthetavsptandcent_PP}}
\end{figure}

Since the initially produced $K^{\ast0}$ has no spin alignment in the UrQMD model, the measurements of all $K^{\ast0}$ with respect to the production plane are similar to those with respect to the reaction plane. As shown in Fig.~\ref{fig:AbsCostheta_RP}(c), the $\cos\theta^*$ distribution is flat for all $K^{\ast0}$ in 20-60\% Au+Au collisions, and the extracted $\rho_{00}=0.3329 \pm 0.0009$ is consistent with 1/3 within uncertainties. The $p_{T}$ dependence is shown in Fig.~\ref{fig:AbsCosthetavsptandcent_PP}(a) as open circles.

\begin{figure}[htbp]
	\centering
	\captionsetup[subfloat]{labelsep=none,format=plain,labelformat=empty}
	\subfloat[]{
		\begin{minipage}[t]{0.45\linewidth}
			\centering
			\includegraphics[width=3in]{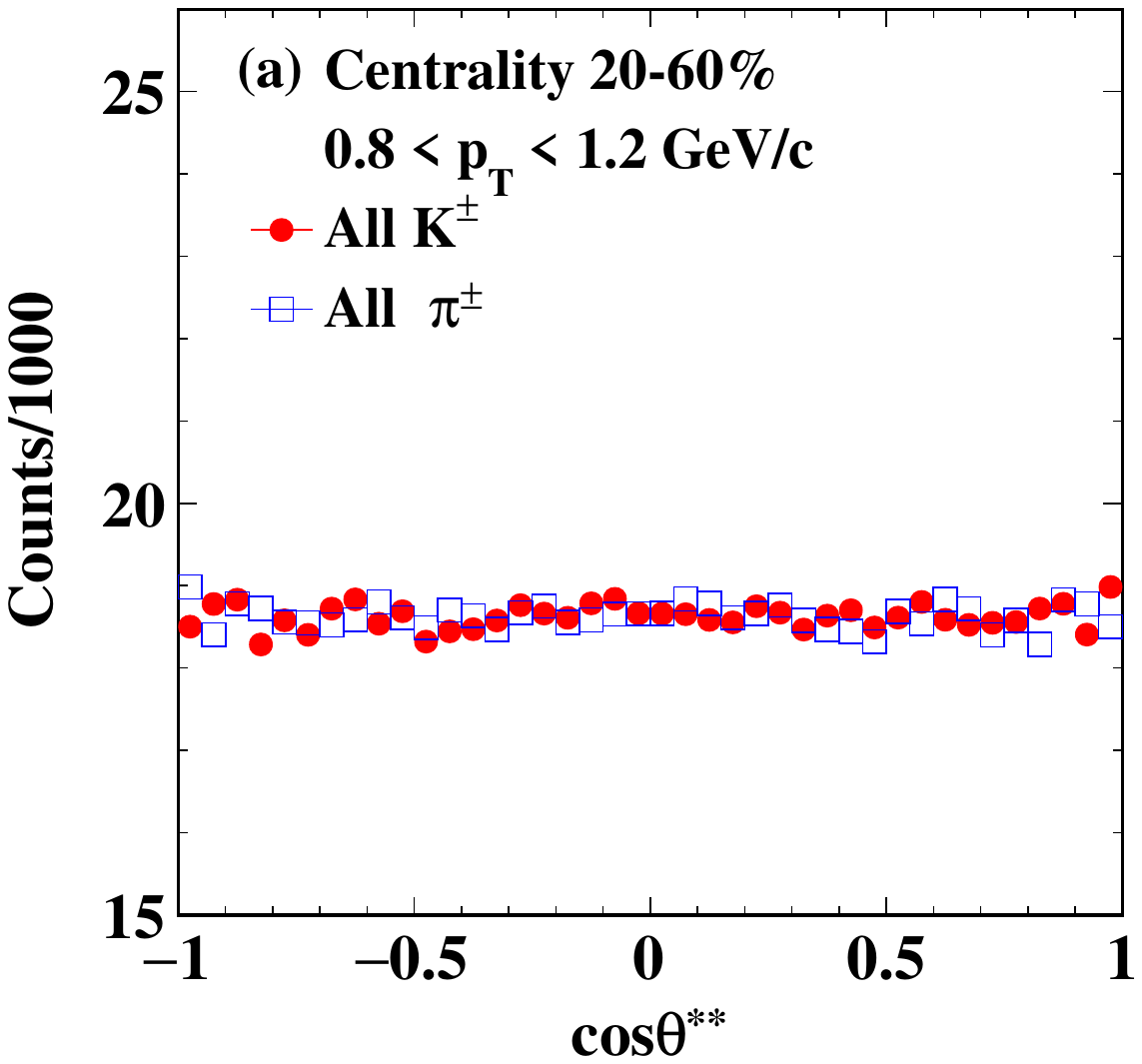}
		\end{minipage}
	}
	\subfloat[]{
		\begin{minipage}[t]{0.45\linewidth}
			\centering
			\includegraphics[width=3in]{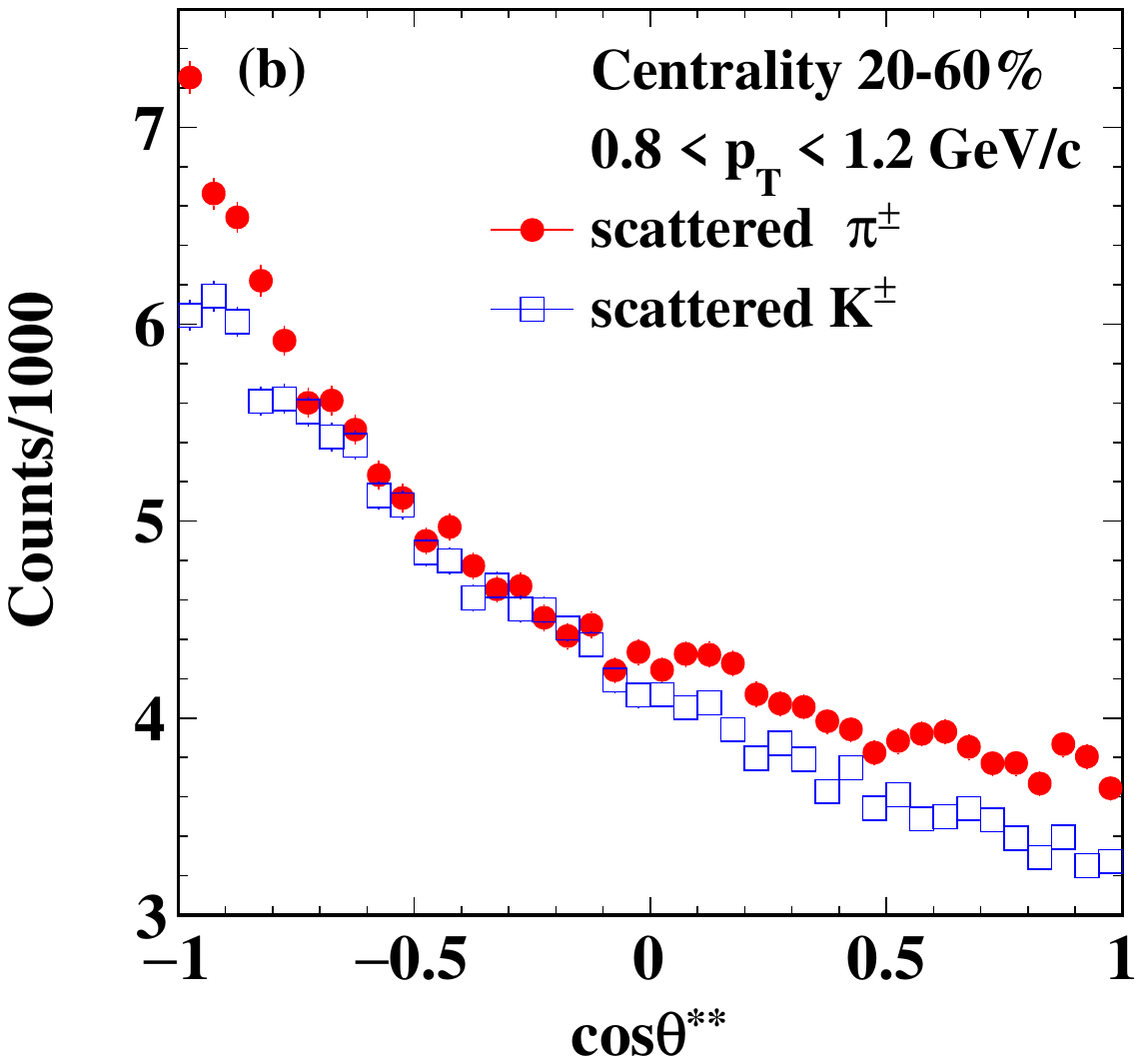}
		\end{minipage}
	}
	
	\centering
	\caption{The $\cos\theta^{**}$ distribution for (a) all decay daughters and (b) scattered decay daughters for 0.8 < $p_{T}(K^{\ast0})$ < 1.2 GeV/c in 20-60$\%$ centrality. 
	\label{fig:Costhetastarstar}}
\end{figure}

We then studied the $\rho_{00}$ value after the rescattering effect with respect to the production plane. Figure~\ref{fig:AbsCostheta_RP}(d) shows that there is less reconstructable $K^{\ast0}$ toward $|\cos\theta^*|=0$ or $\theta^*=\pi/2$, which contradicts with the results with respect to the reaction plane. The extracted $\rho_{00}$ value in this particular $p_{T}$ and centrality bin is $0.3506 \pm 0.0011$, which is larger than 1/3 by 0.017. Figure~\ref{fig:AbsCosthetavsptandcent_PP} shows its $p_{T}$ and centrality dependence. The deviation of reconstructable $K^{\ast0}$ $\rho_{00}$ from 1/3 is more significant in intermediate $p_T$ region and (semi-)central collisions, reaches as large as 0.03 at $0.8<p_T<1.2~\textrm{GeV}/c$ and 0-20\% centrality. This is because the decay daughters are uniformly distributed in the rest frame of $K^{\ast0}$ as we proved before, and daughters that move back to the $K^{\ast0}$ momentum direction in its rest frame will have relatively lower momentum when boosted to the laboratory frame. Then, these daughters are expected to be rescattered more easily since rescattering affects more for low momentum particles. Since the polarization direction is always perpendicular to the $K^{\ast0}$ momentum direction, there are fewer reconstructable $K^{\ast0}$s at $|\cos\theta^*| \approx 0$ than $|\cos\theta^*| \approx 1$. To look at it in more detail, we calculate the distribution of $\theta^{**}$, which is the angle between the momentum direction of the parent $K^{\ast0}$ in the laboratory frame and the momentum direction of a daughter particle in the $K^{\ast0}$ rest frame. Figure~\ref{fig:Costhetastarstar}(a) shows the $\cos\theta^{**}$ distribution for all K or $\pi$, and the distribution is flat. Figure~\ref{fig:Costhetastarstar}(b) shows the $\cos\theta^{**}$ distribution for scattered K and $\pi$, and there are more scattered daughters toward $\cos\theta^{**}=-1$ or in the opposite direction of $K^{\ast0}$. For the results with respect to the reaction plane, as the $K^{\ast0}$ momentum direction is more or less isotropic with respect to the reaction plane, this effect is small.

\section{Summary}\label{sec:summary}
In summary, we have studied the rescattering effect on the spin alignment parameter $\rho_{00}$ of $K^{\ast0}$ in Au+Au collisions at $\sqrt{s_{\mathrm{NN}}}$ = 200 GeV based on the UrQMD model. The results show that the rescattering effect gives a negative contribution to $\rho_{00}$ with respect to the reaction plane. The $p_{T}$ dependence and centrality dependence are also measured. The maximum deviation reaches as large as $-$0.008, which is significant compared with various theoretical sources and comparable to the current experimental precision.The results with respect to the production plane show that the rescattering effect gives a positive contribution to $\rho_{00}$, and the deviation is much larger than that with respect to the reaction plane. This is mainly due to the momentum dependence of the rescattering probability of $K^{\ast0}$ decay daughters. These results indicate that the rescattering effect should be considered in $K^{\ast0}$ spin alignment. 

$Note~added$. Recently, we noticed a study of the same effect using the AMPT model~\cite{Shen:2021pds}. The deviation they found is different from what we find from this study. The difference may come from the different implementations of hadronic rescattering processes in these two models.

\section*{Acknowledgments}
The authors thank Dr. Jinhui Chen, Bedangadas Mohanty, Dr. Aihong Tang, and Dr. Zhangbu Xu for helpful discussions. This work is supported in part by National Natural Science Foundation of China with Grant No. 11720101001, the Strategic Priority Research Program of Chinese Academy of Sciences with Grant No. XDB34030000 and Natural Science Foundation of Anhui Province with Grant No. 2208085J23.
\bibliography{kstar_spin_UrQMD}
\end{document}